# Magnetostructural coupling, magnetic ordering and cobalt spin reorientation in metallic $Pr_{0.50}Sr_{0.50}CoO_3$ cobaltite

José Luis García-Muñoz[1], Jessica Padilla-Pantoja[1], Xavier Torrelles[1], Javier Blasco[2], Javier Herrero-Martín[3], Bernat Bozzo[1], José A. Rodríguez-Velamazán[2,4]

[1]*Institut de Ciència de Materials de Barcelona, ICMAB-CSIC, Campus universitari de Bellaterra, E-08193 Bellaterra, Spain*

[2]*Instituto de Ciencia de Materiales de Aragón, CSIC-Universidad de Zaragoza, 50009 Zaragoza, Spain*

[3] *ALBA Synchrotron Light Facility, 08290 Cerdanyola del Vallès, Barcelona, Spain*

[4] *Institute Laue Langevin, BP 156, 38042 Grenoble Cedex 9, France*

Corresponding author:
José Luis García-Muñoz
Institut de Ciència de Materials de Barcelona,
ICMAB-CSIC
Campus universitari de Bellaterra
08193  Bellaterra, Catalunya, Spain.
e-mail: garcia.munoz@icmab.es

**Abstract**

In half-doped $Pr_{0.50}A_{0.50}CoO_3$ metallic perovskites, the spin-lattice coupling brings about distinct magnetostructural transitions for A=Ca and A=Sr at temperatures close to ~100 K. However, the ground magnetic properties of $Pr_{0.50}Sr_{0.50}CoO_3$ (PSCO) strongly differ from $Pr_{0.50}Ca_{0.50}CoO_3$ ones, where a partial $Pr^{3+}$ to $Pr^{4+}$ valence shift and Co spin transition makes the system insulating below the transition. This work investigates and describes the relationship between the *Imma→I4/mcm* symmetry change [Padilla-Pantoja et al, Inorg. Chem. 53, 12297 (2014)] and the original magnetic behavior of PSCO versus temperature and external magnetic fields. The FM1 and FM2 ferromagnetic phases, above and below the magnetostructural transition ($T_{S1}$~120 K) have been investigated. The FM2 phase of PSCO is composed of [100] FM domains, with magnetic symmetry *Im'm'a* ($m_x\neq 0$, $m_z=0$). The magnetic space group of the FM1 phase is *Fm'm'm* (with $m_x=m_y$). Neutron data analyses in combination with magnetometry and earlier reports results agrees with a reorientation of the magnetization axis by 45º within the *a-b* plane across the transition, in which the system retains its metallic character. The presence below $T_{S1}$ of conjugated magnetic domains, both of *Fm'm'm* symmetry but having perpendicular spin orientations along the diagonals in the xy-plane of the tetragonal unit cell, is at the origin of the anomalies observed in the macroscopic magnetization. A relatively small field $\mu_0 H [\perp z] \gtrsim 30$ mT is able to reorient the magnetization within the *a-b* plane, whereas a higher field ($\mu_0 H [//z] \gtrsim 1.2$ T at 2K) is necessary to align the Co moments perpendicular to *a-b* plane. Such a spin reorientation, in which the orbital and spin components of the Co moment rotate joined by 45º, was not observed with other lanthanides different to praseodymium.

## I. INTRODUCTION

Following the extensive investigations on the nature of the spin-state (SS) changes in undoped $Ln$CoO$_3$ compounds, the SS of trivalent cobalt is being examined in a variety of cobaltites because of its proved ability to condition the transport, magnetic and electronic properties of compounds like the ($Ln_{1-y}Ln'_y$)$_{1-x}A_x$CoO$_3$ ($Ln,Ln'$: lanthanides, $A$: alkaline-earth) perovskites.[1-4] In this context, the physical properties of half-doped Pr-based Pr$_{0.50}A_{0.50}$CoO$_3$ specimens are attracting the interest due to the observation of non-conventional phase transitions and distinct unexpected properties with $A$=Ca [4-11] and Sr.[12-18]

In this way, Ca doped Pr$_{0.50}$Ca$_{0.50}$CoO$_3$ [and other related (Pr,$Ln$)$_{1-x}$Ca$_x$CoO$_3$ cobaltites near half-doping (x~1/2)], exhibits an exotic metal-insulator transition [1] produced by two concurrent phenomena: (i) an abrupt Co$^{3+}$ spin-state change [10] and (ii) a partial Pr$^{3+}$ to Pr$^{4+}$ valence shift.[7-9] Pr$_{0.50}$Ca$_{0.50}$CoO$_3$ (PCCO) is orthorhombic (*Pnma*) and metallic but it becomes insulating at $T_{MI}$ ~ 80 K. An electron is transferred from some Pr atoms to Co sites [6-10] and a concomitant SS crossover promotes the stabilization of the Co$^{3+}$ low spin (LS) state.[10] Remarkably, PCCO exhibits exceptional photoresponse capabilities of potential interest for ultrafast optical switching devices.[11] The generation of metallic domains in the sample after photoirradiation in the non-conducting state occurs thanks to the strong connection between volume expansion, electron mobility and excited spin-states.[10]

The structural, magnetic and electronic properties of Pr$_{0.50}$Sr$_{0.50}$CoO$_3$ (PSCO) apparently differ from PCCO (without magnetic order due to the LS state stabilization in the trivalent Co sites). PSCO is ferromagnetic (FM) below $T_C$ ~ 230 K and metallic in all the temperature range. Mahendiran *et al* [12] initially reported unexpected magnetic anomalies at $T_{S1}$ ~ 120 K. The discovery of a second magnetic transition and intriguing step-like behavior of the magnetization, which decreases or increases depending on the magnitude of the applied field,[12-16] was followed by the detection of structural anomalies at the same temperature by Troyanchuk *et al*.[15] The lack of consensus on the structural properties of PSCO led to different structural descriptions, used to justify visible changes in diffraction data at low temperatures.[13,15-17] On the other hand, some works attributed the magnetostructural transition to a phase separation at T<120 K, proposing a two-phase magnetic state at low temperature.[16] Finally, a reliable description of the crystal structure evolution across $T_{S1}$ was reported in 2014 by Padilla *et al*. in ref. 18. From the high temperature cubic phase, upon decreasing temperature PSCO follows the *Pm-3m* →*R-3c*→*Imma*→*I4/mcm* transformations. Hence the symmetry change at the magnetostructural transition at about $T_{S1}$ implies an orthorhombic-tetragonal conversion (O-T).[18]

The absence of this transition in other half-doped cobaltites without Pr ions and the spontaneous Pr valence shift reported in PCCO and other $(Pr,Ln)_{1-x}Ca_xCoO_3$ cobaltites motivated to investigate the possible importance of the Pr $4f$ – O $2p$ hybridization for the structural changes in PSCO.[13,16,18] Unlike PCCO, a $Pr^{3+}$ to $Pr^{4+}$ oxidation process was ruled out in PSCO by means of x-ray absorption spectroscopy (XAS) studies at Pr $M_{4,5}$ and Pr $L_3$ edges and charge-transfer multiplet calculations.[19] Similarly, XAS measurements of the temperature evolution of the Co $L_{2,3}$ edges showed that the spin state of Co ions remains nearly unaltered across the anomalous transition.[19]

Evidences of the interplay between the magnetic and crystal structures were obtained from transverse susceptibility and magnetostriction measurements which point to likely changes in the magnetocrystalline anisotropy at $T_{S1}$.[12-14,20] Lorentz Transmission Electron Microscopy (LTEM) images reported by Uchida *et al.* also suggested a reorientation of the magnetization axis by 45º when studying the evolution of the magnetic domain structure under electron-beam.[21] The importance of the spin-lattice coupling has been also confirmed by means of x-ray magnetic circular dichroism (XMCD) experiments at the Co $L_{2,3}$ edges.[22] They reveal a sizeable orbital momentum in Co atoms that evolves in like manner as the atomic spin moment (or the macroscopic magnetization) across the two successive magnetic transitions, pointing to a coupling between the ordered electronic spins and the orbital states of $3d$ electrons.

We present a neutron diffraction investigation of the singular magnetic properties of $Pr_{0.50}Sr_{0.50}CoO_3$ that clarifies the temperature and field evolution of the magnetic symmetry in this system. The relevance of the structural symmetry changes on its magnetic behavior has been elucidated.

## II. EXPERIMENTAL DETAILS

Polycrystalline ceramic samples of PSCO were prepared by the conventional solid-state reaction method under an oxygen atmosphere as reported in Ref. 18. High- purity $Co_3O_4$ and $Pr_6O_{11}$ oxides and $SrCO_3$ were used as precursors. The last two annealings were performed at 1100 ºC (for 12h) and 1170 ºC (for 24h) under $O_2$, making a slow cooling. Powder samples and compacted pellets were used for the measurements. Samples quality was checked by x-ray diffraction patterns collected at room temperature using a Siemens D-5000 diffractometer and Cu K$\alpha$ radiation. They were single-phase and free from impurities. The magnetic response to dc and ac magnetic fields was measured using a Superconducting Quantum Interferometer Device (SQUID) and Physical Properties Measuring System (PPMS) from Quantum Design. The latter was also used for electrical transport measurements using the four-probe method and silver paste.

Neutron diffraction experiments were carried at the high-flux reactor of the Institut Laue Langevin (ILL, Grenoble) using the D20 ($\lambda$=1.87 Å) and D1B ($\lambda$=2.52 Å) instruments. Neutron powder diffraction (NPD) measurements in the D20 diffractometer of the ILL were performed using a high take-off angle of 118° for the Ge(115) monochromator, and a radial oscillating collimator that precedes a microstrip PSD detector, covering an angular range of 150°. In this range high resolution data as a function of temperature was obtained warming the sample in a cryofurnace from 15 K up to 443 K. In ramp mode the temperature shift for individual scans was smaller than 5 K. Additional NPD patterns were also recorded at fixed selected temperatures. A cryomagnet was used on D1B to apply magnetic fields up to 5 T. All the structural and magnetic Rietveld refinements were made using the Fullprof program.[23] Crystallographic tools from the Bilbao Crystallographic server were also used.[24-26] A detailed structural study of a PSCO powder sample is reported in ref. 18. Temperature dependent X-ray magnetic circular dichroism measurements at the Co $L_{2,3}$ edges were also performed on the samples at BL29-BOREAS beamline in the ALBA Synchrotron Light Facility, and the results can be found in ref. 22.

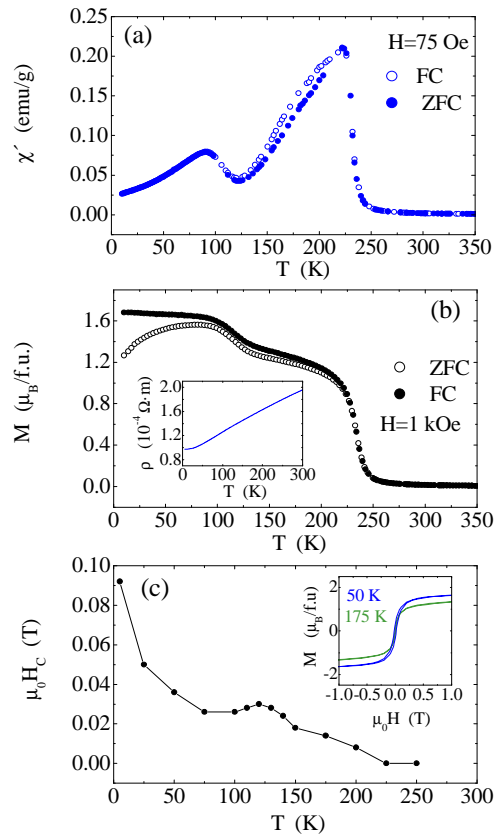

**Figure 1.** (Color online) (a) Real component of ac susceptibility (h=10 Oe, 13 Hz) measured under cooling and heating and a dc field of 75 Oe (FC). (b) FC and ZFC dc magnetization curves (1 kOe). The inset shows the resistivity curve. (c) Temperature evolution of the coercive field ($H_C$). Inset: field dependence of the magnetization (ZFC) at 175 K (FM2) and 50 K (FM1).

## III. RESULTS AND DISCUSSION

The ac susceptibility was measured under a dc field of 75 Oe, superimposed to an ac field of 10 Oe. As illustrated in Figure 1(a), the magnetic transitions produce on cooling pronounced upturns in the real component of the ac susceptibility ($\chi'$), forming two separated peaks with maxima at 92 and 225 K respectively. Regarding the onset of the two peaks in $\chi'(T)$, the first one starts to develop at ~245 K and the second at ~120 K. Magnetization (M) was also measured as a function of temperature and applied magnetic field using a commercial SQUID. A comparison of the zero-field-cooled (ZFC) and field-cooled (FC) magnetizations under 1 KOe plotted in Fig. 1(b) shows a similar hump in both M(T) curves at about $T_{S1}$~120K. A small splitting in the curves is detected below $T_C \approx 230$ K which increases below $T_{S1}$. Moreover, it was reported in earlier works that the sharp decrease in M(T) when cooling under small fields is accompanied by thermal hysteresis.[12,13]

The metallic resistivity is depicted in the inset of Fig. 1(b). The temperature evolution of the coercive field ($H_c$) was determined from the hysteresis loops recorded at diverse temperatures. The obtained $H_c(T)$ curve is represented in Fig. 1(c). Overall, $H_c$ increases with decreasing the temperature but a clear anomaly (peak shape) is observed at $T_{S1}$. In the inset, the field dependence of the magnetization is plotted at two temperatures representative of the two ferromagnetic phases. From now on we will name FM1 to the distinctive ferromagnetic state of the *I4/mcm* phase (T< $T_{S1}$), and FM2 to the ferromagnetic state of the *Imma* cell below the Curie temperature ($T_{S1}$<T< $T_C$). The maximum value of the magnetization M(H) below $T_{S1}$ (at 5 K and 7 T applied field) was very close to 2 $\mu_B$/f.u. (M=1.96$\mu_B$/f.u.).[22]

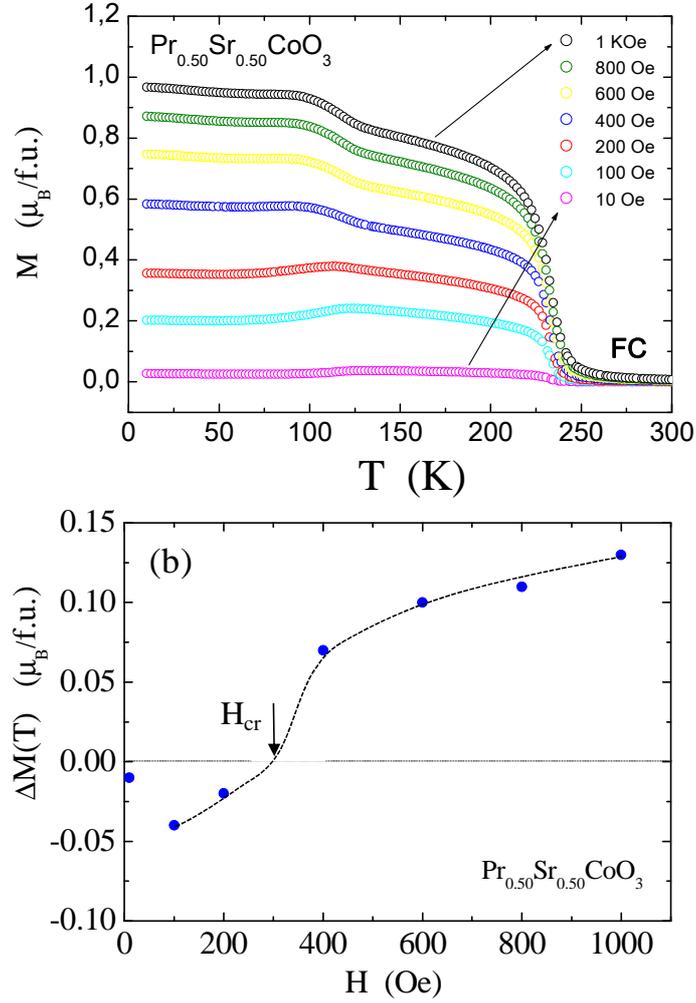

**Figure 2.** (Color online) (a) Temperature dependence of the magnetization measured on cooling at 10, 100, 200, 400, 600, 800 and 1000 Oe (FC). The sign of the magnetization jumps at the second transition are field dependent. (b) Field evolution of the amplitude and sign of the magnetization jumps at the O-T transition.

The temperature dependence of the magnetization measured on cooling under field is shown in Fig. 2(a) for seven distinct magnetic fields. The anomalous step-like behavior of the magnetization around $T_{S1}$ ~ 120 K shows noticeable field dependence. So, the amplitude and sign (positive or negative) of the abrupt magnetization jumps (ΔM) exhibit an ostensive dependence on the strength of the external magnetic field. By inspection of the magnetization jumps shown in the Fig. 2(a) for different fields, the corresponding field dependence has been represented in Fig. 2(b) demonstrating a crossover from negative to positive ΔM values at $T_{S1}$ at the critical value $\mu_0 H_{cr} \approx 30$ mT. The jumps are negative for external fields H < $H_{cr}$, and positive for H > $H_{cr}$.

### III-a Ferromagnetic symmetry in the *Imma* phase (FM2)

Below $T_C \approx 230$ K the emergence of ferromagnetic order produces changes in the neutron profile (ferromagnetic *Imma* phase: FM2). Given the smallness of the orthorhombic distortion it is not easy to unambiguously distinguish the exact orientation of the ferromagnetic Co moments. However, in this case the determination of the moments direction is possible as we will show in this section.

As previously mentioned, a detailed description of the crystal structures of PSCO can be found in ref. 18. Let us recall first -to avoid possible confusions- that the standard settings of the space groups (SGs) *Imma* and *I4/mcm* correspond to different orientations of the perovskite cell: $\sqrt{2}a_0$ x $2a_0$ x $\sqrt{2}a_0$ in the orthorhombic SG and $\sqrt{2}a_0$ x $\sqrt{2}a_0$ x $2a_0$ in the tetragonal one. Namely, the longest cell parameter is *b* in the first setting and *c* in the second one. To avoid misunderstandings, hereafter we will label the ferromagnetic models $F_{x,y \text{ or } z}$ always referred to the unit cell setting of the tetragonal phase, where z denotes the coordinate along the longest axis $2a_0$. Therefore, under this definition, $F_z$ along this paper always means that spins are pointing parallel to the longest (vertical) $2a_0$-axis, which is the *c* axis in *I4/mcm* but it is the *b-one* in *Imma* phase.

In an effort to obtain some insight on the anisotropy of the magnetization in the orthorhombic phase, we performed careful refinements of the D20 neutron diffraction pattern at 140 K using different moment orientations. The differences using different magnetic anisotropies were small. Nevertheless the best fit was obtained for the $F_x$ configuration. $F_y$ or $F_z$ models generate wrong ferromagnetic intensities (in comparison to the experimental ones) in well resolved peaks like (123) and (321), being experimentally higher the magnetic contribution to the last. Moreover, these two models also produce an excess of (103) magnetic intensity, whereas the $F_x$ configuration nicely reproduces all magnetic intensities.

In addition, several possible magnetic or Shubnikov space groups (MSGs) compatible with the *Imma* symmetry and **k**=0 were considered. Among them the magnetic subgroups *Im'm'a* (allowing $F_xG_z$, expressed in the tetragonal setting) or *Imm'a'* (compatible with $F_y$ and $G_xF_z$, referred to the tetragonal setting).[26,27] The best fit was obtained with the MSG *Im'm'a* [# 74.558, transformation to standard setting:(**a**, **b**, **c**; 0, 0, 0)] [25,27], compatible with a collinear $F_x$ model (tetragonal setting), which yields goodness factors $R_B$=3.38, $R_f$=2.74, $R_{Mag}$=4.89, $\chi^2$=1.58 (see details in Table I). The Rietveld refinement at 140 K (above but very close to the O-T phase transition) using the *Im'm'a* MSG symmetry converges to the $F_x$ model (and $m_z$=0, tetragonal setting) and it is plotted in Figure 3, yielding a FM ordered moment $m_x$=1.49(3) $\mu_B$/Co. Schematic projections of the magnetic order in the orthorhombic cell are shown in the Table I and Fig. 3. The cell parameters and atomic coordinates coincide with those reported in ref. 18. So, $a$=5.3771(8) Å, $b$=7.5950(1) Å, $c$=5.4320 (7) Å; Pr/Sr (4$e$): $z$[Pr/Sr]= -0.0005(4); Co (4$b$); O1 (4$e$): $z$[O1]= 0.4544(6); O2 (8$g$): $y$[O2]= 0.0243(2).

Likewise, we confirmed that lower magnetic symmetries allowing ferromagnetic order out of the main crystallographic axes (either in the *x-y* plane or out of this plane) produce worse results and generate wrong magnetic intensities in reflections like (101), (020), (103), (301) or (123). As example, the subgroup (of *Im'm'a*) *C2'/m'* allows the FM spins to collectively rotate within the *x-y* plane ($F_xF_y$ model, expressed in the tetragonal setting). Using this MSG (forcing identical moments in the split Co orbits) the refinements converge to moments aligned along *x*, with negligible *y* component (tetragonal setting). Summarizing then, we have been able to discern the x-orientation of the magnetic moments in the FM2 ferromagnetic phase (MSG *Im'm'a*, $m_x\neq0$) of the orthorhombic *Imma* structure. Paying attention to fine details in the magnetic Rietveld refinement when comparing different models we conclude that only the $F_X$ model (tetragonal setting) correctly matches the experimental intensities.

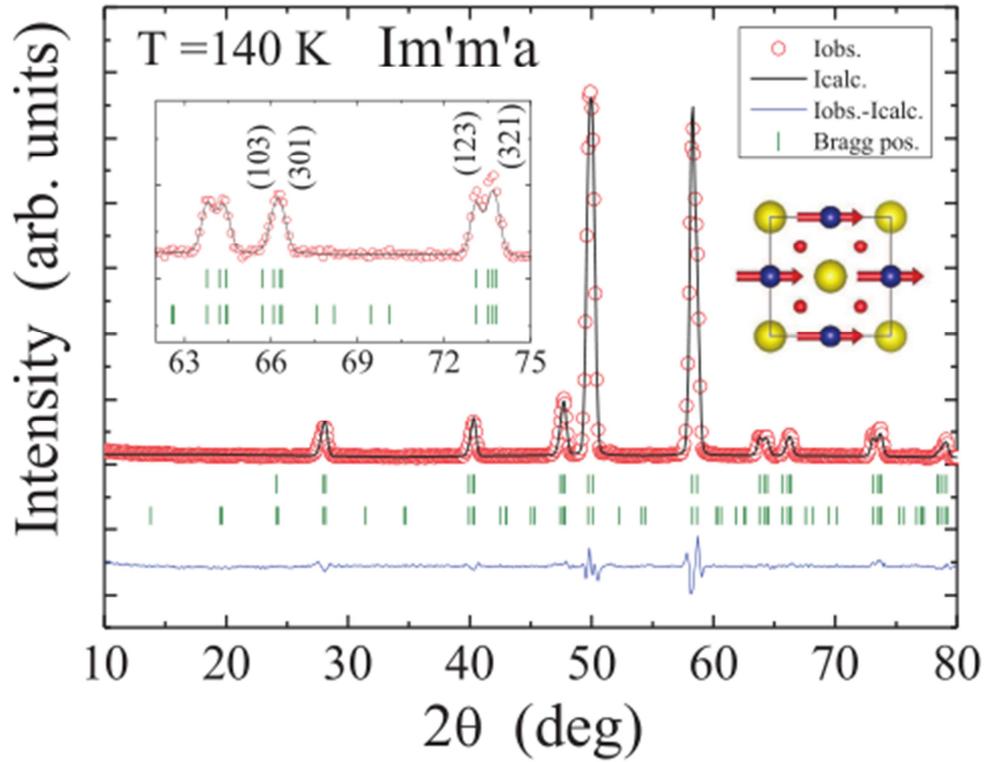

**Figure 3.** (Color online) Rietveld refinement (solid line) of neutron pattern collected at 140 K (FM2 phase, D20) for the ferromagnetic orthorhombic phase *Im'm'a*. The inset shows indexed magnetic peaks, $\sqrt{2}a_0 \times 2a_0 \times \sqrt{2}a_0$ setting. An schematic view of the magnetic structure is also shown ( yellow balls are Pr/Sr atoms, blue are Co atoms and O atoms are red).

**III-b Magnetic symmetry and cobalt spin reorientation in the *I4/mcm* phase (FM1)**

Figure 4(left panel) plots the evolution of two main sets of magnetic reflections: (110)/(002) and (112)/(020) at 2θ nearby 28º and 40º, respectively (D20 data, λ=1.87 Å). The two transitions can be easily identified. The intensities shown in the figure were recorded in absence of external magnetic field, and their evolution show a clear decrease at the magnetostructural transition which reminds the step-like behavior of M(T) measured at low fields. However, the changes observed in the neutron patterns around 120 K may arise from potentially concurrent structural and magnetic intensity changes. Thereby, we will show later that the loss of intensity in the left panel curves of Fig. 4 is not due to a decrease of the FM Co moment (as changes in magnetization could suggest) but it simply has a structural origin. Therefore, Fig. 4 is just an illustrative example that the concurrent structural and magnetic changes in the new tetragonal cell favor cross-correlation between these two types of parameters. For that reason we decided to firstly explore possible magnetic orders compatible with the new tetragonal symmetry, and confront them to neutron data. Neutron patterns in the low-temperature phase do not show new magnetic reflections symptomatic of an antiferromagnetic (AFM) multiple cell (**k** is 0). Nevertheless, the possibility of canted Co moments with possible coexisting FM and AFM coupled components in the tetragonal cell phase was considered too. Among the most probable magnetic symmetries compatible with the *I4/mcm* cell in absence of external field one can include the maximal subgroups: *I4/mc'm'* (#140.547), *Iba'm'* or *Ib'am'* (#72.544) and *Fm'm'm* (#69.524),[27] all of them compatible with a net ferromagnetic behavior (collinear or non-collinear) from cobalt atoms sharing one single orbit.[24-26] Some other subgroups of the grey magnetic group *G1'* (*G= I4/mcm*) were ignored because they are not compatible with ordered moments at Co sites (like *Iba'm* or *Ib'am*) or imply pure antiferromagnetism [like *I4/mcm* (#140.541)].

*III-b-1 Discerning between out-of-plane and in-plane ordering*

The tetragonal magnetic symmetry *I4/mc'm'* (allowing only $m_z$ components, $F_z$ model) was immediately discarded as it clearly generates wrong magnetic intensities. It is shown in Fig. 4(right panel) that considering Co moments aligned along [001] the experimental magnetic intensities from the (00l) l=2n planes cannot be reproduced. Hence, unlike the ferromagnetic I4/mcm phase of the manganite $Pr_{0.50}Sr_{0.50}MnO_3$ where the magnetization vector lies along [001],[28] the magnetic moments in the cobalt counterpart compound -$Pr_{0.50}Sr_{0.50}CoO_3$- are within the *x-y* plane of the tetragonal cell ($F_{\perp z}$).

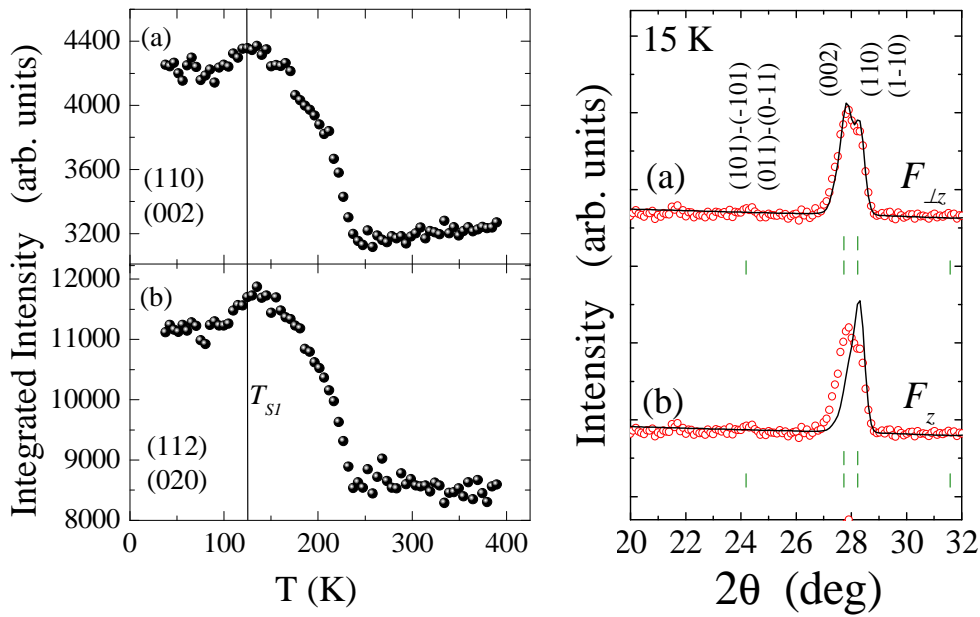

**Figure 4.** (Color online) (Left panel) Evolution of the integrated intensity of two main sets of diffracted reflections with ferromagnetic contribution: (a) (110)/(002) and (b) (112)/(200). Neutron diffraction measurement heating the sample without external field. (Right panel) Selected region of the refined neutron pattern at 15 K using different orientations for the Co magnetic moments: (a) FM model with in-plane magnetic moments ($F_{\perp z}$ model) and (b) FM model with out-of-plane magnetic moments ($F_z$ model). The reflections are indexed in the *I4/mcm* setting ($\sqrt{2}a_0 \times \sqrt{2}a_0 \times 2a_0$ cell).

*III-b-2 Making compatible neutron and LTEM results*

As the best solution for the magnetic order in the tetragonal phase we found the *Fm'm'm* (#69.524) magnetic SG [transformation to standard setting:(**-c, a-b, -a-b** ; 0, 1/2, 0)] with Co having $m_x=m_y=1.32(3)$ $\mu_B$/atom as refined values, resulting in Co moments pointing along the diagonal [110] direction within the *a-b* plane of the cell (see Table I). The refinement of the 15 K pattern with this symmetry is perfectly satisfactory, yielding a FM ordered moment of 1.87(4) $\mu_B$/Co. Any other tested model did not show better reliability factors: $R_B$=4.81, $R_f$=3.16, $R_{Mag}$=8.55, $\chi^2$=1.59. Of course, as a matter of fact, the rotation of this magnetic structure around the tetragonal axis does not modify the quality of the neutron refinement. *Iba'm'* and *Ib'am'* magnetic symmetries correspond to the conjugated magnetic domains associated, respectively, to FxGy and GxFy models (#72.544), permitting FM order along *x* or *y* directions (equivalent in tetragonal symmetry). Nevertheless the magnetic models with the magnetization vector parallel to the x or y axis of the tetragonal cell ($\sqrt{2}a_0$) can be discarded because they are not compatible with the low temperature observations made on PSCO by Lorentz transmission electron microscopy reported in ref. 21, which imply a rotation of the moments by 45º across the transition. We shall return to this point later. The measured, calculated and difference profiles for tetragonal PSCO are plotted in Figure 5. A close view of some refined important magnetic reflections is presented in the inset.

As shown in Table I, the *Fm'm'm* magnetic symmetry permits two independent components ($m_x$ and $m_y$, expressed in the tetragonal setting) perpendicular to the vertical tetragonal axis. If $|m_x|\neq|m_y|$ the diagonal ferromagnetic FM[110] order is splitted into two non-collinear sublattices that deviate from the diagonal line. Calling $\delta$ to the angle formed by the Co moments at (0,0,0 | $m_x,m_y$,0) and (1/2,1/2,0 |$m_y,m_x$,0) tetragonal sites, it is important to emphasize that the intensity of (101)/(110) reflections would be proportional to $\delta$ if the moments were not collinear. Collinearity is here experimentally confirmed because these reflections have null intensity at $2\theta$=24.53º in Fig. 5, leading to the diagonal $m_x=m_y$ model ($F_{xy}$).

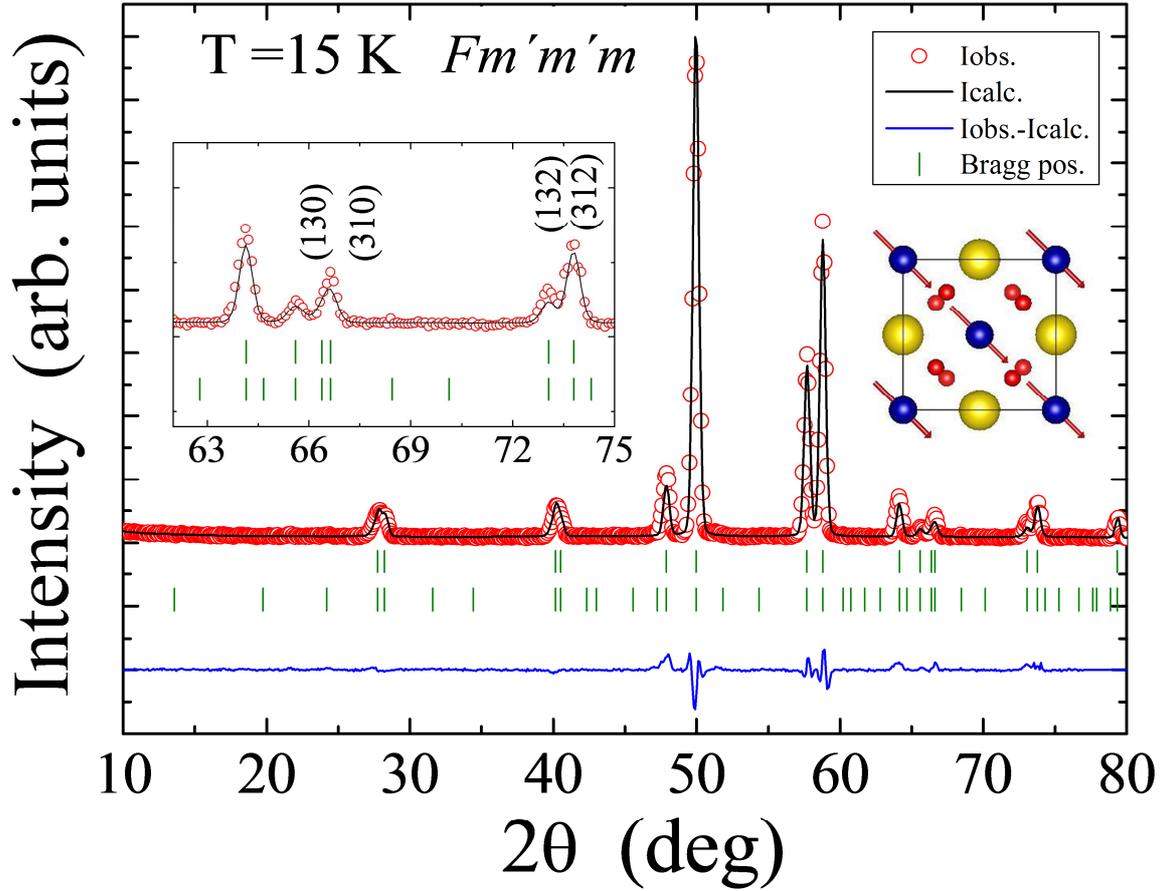

**Figure 5.** (Color online) Rietveld refinement (solid line) of the neutron pattern collected at 15 K (FM1 phase, D20) for the *I4/mcm* structure using the *Fm´m´m* magnetic symmetry with $m_x = m_y$ ($F_{xy}$). The inset shows indexed magnetic peaks, $\sqrt{2}a_0 \times \sqrt{2}a_0 \times 2a_0$ setting. A schematic view of the magnetic structure is also shown.

### III-c Temperature evolution of the ordered FM moment (zero field NPD)

The amplitude of the ordered FM moment per cobalt atom was refined as a function of temperature using the set of neutron patterns collected below $T_c$. Its evolution is shown in Figure 6(a), using the *Im´m´a* (F[100]) and *Fm'm'm* (F[110]) magnetic models for, respectively, the orthorhombic FM2 and tetragonal FM1 phases. In this evolution there is a small partial disruption of the ordered moment around $T_{S1}$ (between 100 K ≤ T ≤ 130 K), attributed to the natural disorder across the O-T structural transition. Anyway, we do not observe the characteristic step down of the magnetization under very low fields. These results confirm the different evolution of the ordered atomic magnetic moment respect to the magnetization behavior.

The samples here presented were previously investigated by x-ray magnetic circular dichroism (XMCD) measurements at the Co $L_{2,3}$ edges,[22] which revealed an unquenched orbital angular momentum in Co atoms. In Fig. 6(b) we reproduce (as adapted from XMCD data in ref. 22) the temperature dependence of the orbital magnetic moment $m_L$ under 0.1 T. The $m_L(T)$ evolution showing an upturn in the figure contrasts with the typical behavior of the ordered ferromagnetic moment deduced from neutrons and depicted in Fig. 6(a).

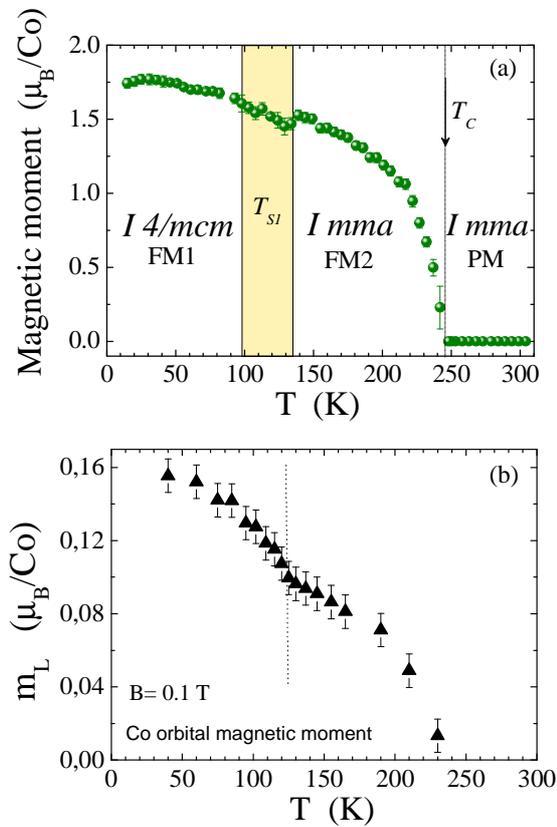

**Figure 6.** (Color online) (a) Temperature dependence of the ordered ferromagnetic moment per Co atom across the Curie and *Imma* to *I4/mcm* transitions obtained from the Rietveld refinement of neutron patterns. (b) Temperature dependence of the orbital magnetic moment $m_L$ (solid blue triangles) derived from XMCD spectra of PSCO at 0.1T (adapted from ref. 22).

### III-d Neutron powder diffraction under magnetic field at 2K

Neutron diffraction experiments were extended to measurements under magnetic field in the low temperature FM1 phase. For that a cryomagnet was placed on the D1B powder difractometer. A sintered cylindrical bar of PSCO was ZFC down to 2K. Then isothermal NPD measurements under field were carried out while increasing the vertically applied magnetic field up to a maximum value of 5T. NPD patterns were collected under constant fields within the 0→1 T ($\mu_0 \Delta H$= 0.1 T step) and 1→5 T (0.5 T step) intervals.
.

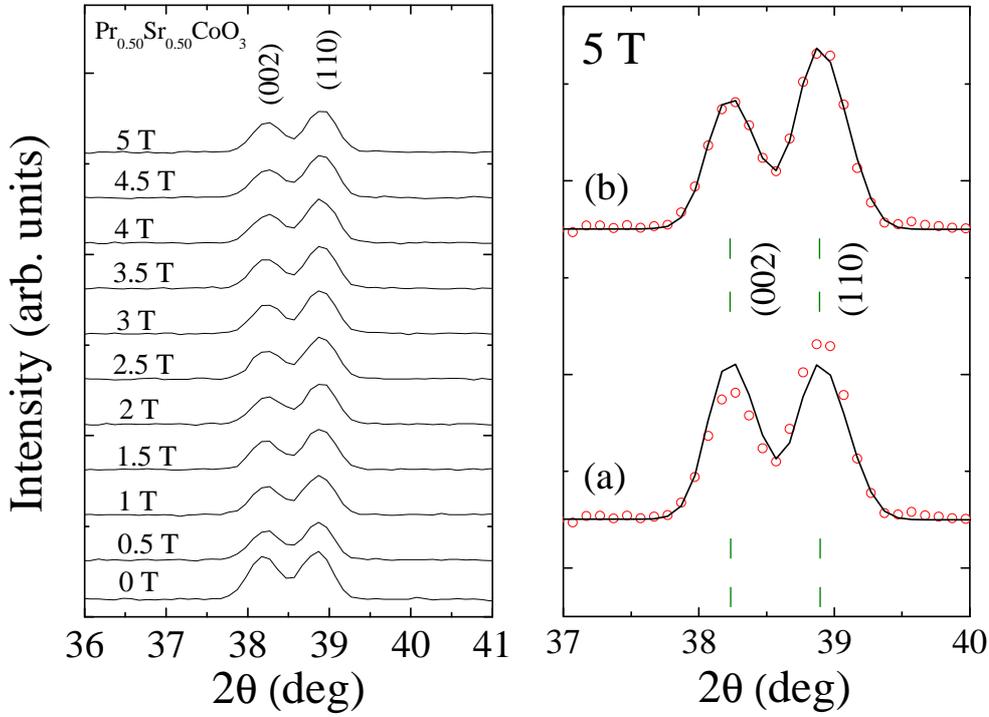

**Figure 7.** (Color online) PSCO, neutron powder diffraction at 2K ($\lambda$=2.52 Å). [Left panel] Evolution of the (002) and (110) reflections increasing the applied magnetic field (see explanation in the text). [Right panel] Selected angular region of the Rietveld refinements (solid line) at 2K under 5T corresponding to: (a) $F_{\perp z}$ model (in-plane ferromagnetism) and (b) In-plane ($F_{\perp z}$) + out-of-plane ($F_z$) model with inclined FM moments deviated 28(2)° from the plane. Patterns have been shifted up for clarity and are indexed in the tetragonal cell.

Figure 7(left panel) displays the evolution upon increasing the magnetic field of the (002)/(110) magnetic peaks at 2 K. By simple inspection one can see that the intensity of both reflections is very similar for H=0, whereas the applied field generates a visible increase of the (110) magnetic intensity with respect to the (002). Even so, the changes in the spectra are rather small. We confirmed that the application of moderate magnetic fields does not destabilize the *I4/mcm* structure. Following these observations, the neutron pattern obtained at 5 T, was satisfactorily refined using the tetragonal crystal structure and FM Co moments with a component within the *a-b* plane ($F_{\perp z}$) and an additional out-of-plane component ($F_z$). The right panel of the Fig. 7 plots the (002)/(110) region of the Rietveld refinement of neutron data under 5 T at 2K (a) constraining the FM moments to the *xy* plane, and (b) permitting also an out-of-plane component in the magnetization ($F_{xy}F_z$). Only this last model does correctly account for the experimental profile, and the best magnetic refinement under 5T yields $m_{Co}[5T]$=2.00(2) $\mu_B$/Co and $\theta$=28(2)° ($R_B$=2.71, $R_f$=1.64, $R_{Mag}$=1.95, $\chi^2$=1.36), where the angle $\theta$ defines the average deviation of the FM moments in the polycrystalline bar out of the *a-b* plane. Using the same model we have satisfactorily refined the successive neutron patterns recorded at fields between 0 and 5 T (2 K). The two refined magnetic parameters (total atomic magnetic moment and the deviation angle $\theta$ out of the *a-b* plane) are depicted in Figure 8 as a function of the external magnetic field. One observes that, upon increasing the applied field from zero, the inclination of the moment (angle $\theta$) first increases but then it saturates reaching a constant value between 1.0 and 1.5 T. So, the evolution of our polycrystalline sample shown in this figure indicates that an external field $\mu_0H_Z$ ~1.2 T applied to a single crystal in the FM2 phase would be enough to align all Co moments perpendicular to the plane (with $F_z$ configuration).

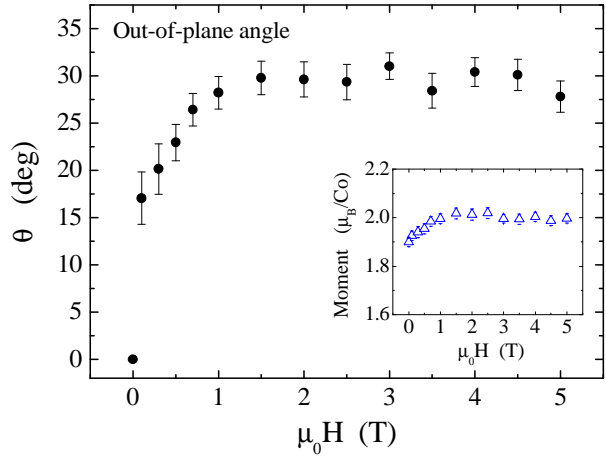

**Figure 8.** (Color online) Evolution upon increasing the magnetic field of the out-of-plane FM component in a polycrystalline bar of PSCO (θ: inclination angle). Inset: ordered FM moment per Co ion.

## IV. CONCLUDING REMARKS

In the precedent sections we have investigated the puzzling magnetic properties of the half-doped PSCO cobaltite. Even though the structural evolution with temperature was previously and extensively described in ref. 18, its magnetic properties and the nature of the magnetostructural transition were not well understood. An orbital contribution to the magnetization (around 1/3 of the spin component) was revealed by XMCD, and the spin-orbit coupling term in this cobaltite promotes a parallel alignment of spin and orbital magnetic moments at both sides of the magnetostructural transition.[22] Very likely a reorientation of the orbital moment at $T_{S1}$ is triggered by the *Imma* → *I4/mcm* phase transition and, through the spin-orbit coupling, the ordered Co spins rotate between the FM1 and FM2 phases. Indeed, two early reports suggested a rotation by 45º of the magnetic easy-axis: Hirahara *et al.*[20] reported preliminary magnetization measurements on a PSCO single crystal (described as monoclinic and only partially characterized); equally, Lorentz Transmission Electron Microscopy images obtained by Uchida *et al.*[21]

are concurrent with a 45° rotation of the magnetization between 140 K and 80 K and with our neutron data analysis.

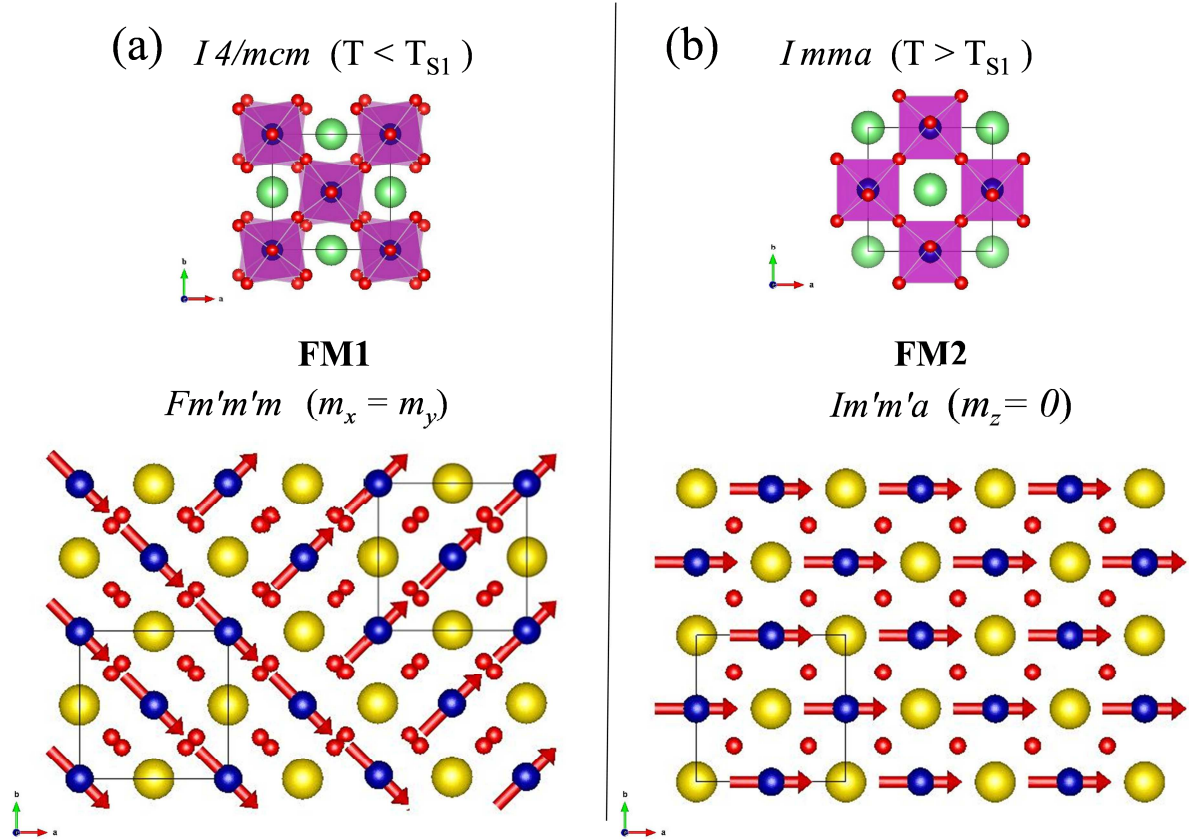

**Figure 9.** (Color online) Schematic view of the crystal and magnetic structures for (right) the ferromagnetic phase above $T_{S1}$ (FM2) composed of [100] ferromagnetic domains and (left) the ferromagnetic phase below $T_{S1}$ (FM1) showing the coexistence of conjugated [110] and [1-10] magnetic domains.

Together with previous LTEM results, the analysis of the magnetic structures using neutron diffraction confirms a spin reorientation accompanying the higher symmetry of the tetragonal *I4/mcm* cell. We have demonstrated that above $T_{S1}$ the FM2 phase of PSCO is composed of [100] FM domains ($F_x$), with magnetic symmetry *Im'm'a* ($m_x \neq 0$, $m_z = 0$). For clarity reasons, the coordinates *x-y-z* are always referred to the setting of the tetragonal cell ($\sqrt{2}a_0 \times \sqrt{2}a_0 \times 2a_0$). Below $T_{S1}$ there is a change in the magnetocrystalline axis. The coupled orbital and spin components of the moment rotate by 45° and the easy axis aligns parallel to the diagonal of the tetragonal unit cell ($F_{xy}$). The magnetic space group of the low temperature phase is *Fm'm'm* (with $m_x=m_y$). The appearance of the 4-fold tetragonal axis brings on degeneration and two equivalent magnetic easy-axes, which are associated to the two conjugated magnetic domains with spin orientations [110] and [1-10]. Therefore, the loss of magnetization (negative step) under low fields (H< $H_{cr}$) is produced by the presence of conjugated [110] and [1-10] ferromagnetic domains after the *Imma* → *I4/mcm* transition. A schematic view of magnetic ordering and magnetic domains above and below $T_{S1}$ is shown in Figure 9. For clarity reasons, time-reversal type domains are not shown in this figure. The coexistence of different types of domains at low temperatures depends on the sample history and the external applied field. On the other hand, the origin of the sudden positive jump in the magnetization under moderate fields could be ascribed to a larger $J/K_a$ ratio, with $J$ and $K_a$ being the double-exchange term and the magnetocrystalline energy, respectively. In the tetragonal phase the Co-O-Co angle parallel to *c* becomes completely flat and could favor the ferromagnetic double-exchange.[18]

Recapitulating, the evolution of neutron diffraction data through the two successive magnetic transitions in $Pr_{0.50}Sr_{0.50}CoO_3$ agrees with a reorientation of the easy-axis of cobalt atoms favored by the *Imma* → *I4/mcm* symmetry change at $T_{S1}$=120 K. The magnetic symmetry *Im'm'a* ($m_x\neq 0$, $m_z=0$) in the orthorhombic phase below $T_C$ generates [100] type ferromagnetic domains ($F_x$). Neutron data analyses in combination with earlier reports agrees with a reorientation of the magnetization axis by 45º within the *a-b* plane in the low temperature phase ($F_{xy}$). The presence below $T_{S1}$ of conjugated magnetic domains of *Fm'm'm* symmetry with spin orientations [110] and [1-10] ($|m_x|=|m_y|$) is at the origin of the anomalies observed in the macroscopic magnetization. A relatively small field of $\mu_0 H_{cr}$ [⊥z] ≈ 30 mT is able to reorient the magnetization within the *a-b* plane, whereas a higher field ($\mu_0 H_{cr,[//z]} \sim 1.2$ T at 2K) would be necessary to align the Co moments perpendicular to this plane.

**ACKNOWLEDGMENTS**

We thank financial support from the Spanish MINECO under projects MAT2012-38213-C02-02 and MAT2012-38213-C02-01 (cofunded by ERDF from EU) and CSD2007-00041 (NANOSELECT). We also acknowledge ILL and ALBA for granting beamtime. C. Ritter is acknowledged for technical assistance during neutron measurements. J.L.G-M thanks J. M. Pérez-Mato for fruitful discussions. J.P-P and J.A.R.V. thank CSIC for JAE-Pre and JAE-Doc contracts, respectively.

**REFERENCES**

[1] S. Tsubouchi, T. Kyômen, M. Itoh, P. Ganguly, M. Oguni, Y. Shimojo, Y. Morii, and Y. Ishii, Phys. Rev. B **66**, 052418 (2002).

[2] C. Frontera, J. L. García-Muñoz, A. Llobet, and M. A. G.Aranda, Phys. Rev. B **65**, 180405(R) (2002). Also C. Frontera, J.L. García-Muñoz, A. E. Carrillo, M. A. G. Aranda, I. Margiolaki, and A. Caneiro, J. Sol. Stat. Chem. 171, 349-352 (2003).

[3] A. Maignan, V. Caignaert, B. Raveau, D. Khomskii, G. Sawtzky, Phys. Rev. Lett., **93**, 026401 (2004). Also C. Frontera, J.L. García-Muñoz, A. E. Carrillo, M. A. G. Aranda, I. Margiolaki, and A. Caneiro, Phys. Rev. B **74**, 054406 (2006).

[4] J. Herrero-Martín, J. L. García-Muñoz, K. Kvashnina, E. Gallo, G. Subías, J. A. Alonso, and A. J. Barón-González, Phys. Rev. B **86**, 125106 (2012).

[5] A. J. Barón-González, C. Frontera, J. L. García-Muñoz, J. Blasco, and C. Ritter, Phys. Rev. B **81**, 054427 (2010).

[6] J. Hejtmánek, E. Šantavá, K. Knížek, M. Maryško, Z. Jirák, T. Naito, H. Sasaki, and H. Fujishiro, Phys. Rev. B **82**, 165107 (2010).

[7] K. Knizek, J. Hetjmánek, P. Novák, and Z. Jirák, Phys. Rev. B **81**, 155113 (2010)

[8] J. L. García-Muñoz, C. Frontera, A. J. Barón-González, S. Valencia, J. Blasco, R. Feyerherm, E. Dudzik, R. Abrudan, and F. Radu, Phys. Rev. B **84**, 045104 (2011)

[9] J. Herrero-Martín, J. L. García-Muñoz, S. Valencia, C. Frontera, J. Blasco, A. J. Barón-González, G. Subías, R. Abrudan, F. Radu, E. Dudzik, and R. Feyerherm, Phys. Rev. B **84**, 115131 (2011).

[10] J. Herrero-Martín, J. L. García-Muñoz, K. Kvashnina, E. Gallo, G. Subías, J. A. Alonso, and A. J. Barón-González, Phys. Rev. B **86**, 125106 (2012)

[11] Y. Okimoto, X. Peng, M. Tamura, T. Morita, K. Onda, T. Ishikawa, S. Koshihara, N. Todoroki, T. Kyomen, and M. Itoh, Phys. Rev. Lett. **103**, 027402 (2009).

[12] R. Mahendiran and P. Schiffer, Phys. Rev. B **68**, 024427 (2003).

[13] C. Leighton, D. D. Stauffer, Q. Huang, Y. Ren, S. El-Khatib, M. A. Torija, J. Wu, J. W. Lynn, L. Wang, N. A. Frey, H. Srikanth, J. E. Davies, Kai Liu, and J. F. Mitchell, Phys. Rev. B **79**, 214420 (2009).

[14] N. A. Frey Huls, N. S. Bingham, M. H. Phan, H. Srikanth, D. D. Stauffer and C. Leighton, Phys. Rev. B **83**, 024406 (2011).

[15] I. O. Troyanchuk, D. V. Karpinskii, A. N. Chobot, D. G. Voitsekhovich, and V. M. Bobryanskii, JETP Lett. **84**, 151 (2006).

[16] A. M. Balagurov, I. A. Bobrikov, V. Y. Pomjakushin, E. V. Pomjakushina, D. V. Sheptyakov and I. O. Troyanchuk, JETP Lett. **93**, 263 (2011). Also A. M. Balagurov, I. A. Bobrikov, D. V. Karpinsky, I. O. Troyanchuk, V. Y. Pomjakushin, and D. V. Sheptyakov, JETP Lett. **88**, 531 (2008).

[17] F. Li, N. Wu and J. Fang, J. Supercond. Nov. Magn 26, 463 (2013).

[18] J. Padilla-Pantoja, J. L. García-Muñoz, B. Bozzo, Z. Jirák and J. Herrero-Martín, Inorg. Chem. **53**, 12297 (2014). See also Additions and Corrections in the related article: Inorg. Chem. **54** (12), 6062 (2015).

[19] J. Padilla-Pantoja, J. Herrero-Martín, P. Gargiani, M. Valvidares, V. Cuartero, K. Kummer, O. Watson, N. Brookes and J. L. García-Muñoz, Inorg. Chem. **53**, 8854 (2014)

[20] S. Hirahara, Y. Nakai, K. Miyoshi, K. Fujiwara, and J. Takeuchi, J. Magn, Magn. Mat, **310**, 1866 (2006).

[21] S. Uchida, R. Mahendiran, Y. Tomioka, Y. Matsui, K. Ishizuka, and Y. Tokura, Appl. Phys. Lett. **86**, 131913 (2005).

[22] J. Padilla-Pantoja, J. Herrero-Martín, E. Pellegrin, P. Gargiani, S. M. Valvidares, A. Barla and J. L. García-Muñoz, Phys. Rev. B **92**, 245136 (2015)

[23] J. Rodríguez-Carvajal, Physica B 192, 55 (1993); [http://www.ill.eu/sites/fullprof/]

[24] J. M. I. Aroyo, J. M. Perez-Mato, C. Capillas, E. Kroumova, S. Ivantchev, G. Madariaga, A. Kirov and H. Wondratschek. "Bilbao Crystallographic Server I: Databases and crystallographic computing programs". Z. Krist. **221**, 1, 15-27 (2006). (http://www.cryst.ehu.es).

[25] M. I. Aroyo, A. Kirov, C. Capillas, J. M. Perez-Mato and H. Wondratschek. "Bilbao Crystallographic Server II: Representations of crystallographic point groups and space groups". Acta Cryst. A62, 115-128 (2006).

[26] J.M. Perez-Mato, S.V. Gallego, E.S. Tasci, L. Elcoro, G. de la Flor, and M.I. Aroyo "Symmetry-Based Computational Tools for Magnetic Crystallography". Annu. Rev. Mater. Res. **45**, 217-248 (2015).

[27] The magnetic space groups are referred to the Belov-Neronova-Smirnova (BNS) notation. N.V. Belov, N.N. Neronova, T.S. Smirnova, *Sov. Phys. Crystallogr.* **2**(2), 311 (1957).

[28] F. Damay, C. Martin, M. Hervieu, A. Maignan, B. Raveau, G. Andre, F. Boure, J. Mag. Mag. Mat. 184, 71-82 (1998).

**Table I.** Magnetic groups and refined magnetic moments in the FM1 (15 K) and FM2 (140 K) phases. See explanation in the text.

| Magnetic Space Group | *Fm'm'm* (#69.524) | *Im'm'a* (#74.558) |
|---|---|---|
| Transformation to standard setting | (-**c**, **a-b**, -**a-b** ; 0, 1/2, 0) | (**a**, **b**, **c**; 0, 0, 0) |
| Coordinates for Co moment | Expressed in parent Tet. setting:<br>(0,0,0 \| $m_x, m_y, 0$)<br>(0,0,1/2 \| $m_y, m_x, 0$)<br>(1/2,1/2,1/2 \| $m_x, m_y, 0$)<br>(1/2,1/2,0 \| $m_y, m_x, 0$) | Expressed in parent Ort. setting:<br>(0,0,1/2 \| 0, $m_y, m_z$)<br>(0,1/2,1/2 \| 0, $-m_y, m_z$)<br>(1/2,1/2,0 \| 0, $m_y, m_z$)<br>(1/2,0,0 \| 0, $-m_y, m_z$)<br><br>Expressed in Tet. setting:<br>(0,0,0 \| $m_x$, 0, $m_z$)<br>(0,0,1/2 \| $m_x$, 0, $-m_z$)<br>(1/2,1/2,1/2 \| $m_x$, 0, $m_z$)<br>(1/2,1/2,0 \| $m_x$, 0, $-m_z$) |
| Refined moments (*expressed in Tet. setting*) | $m_x = m_y = 1.32(3)$ $\mu_B$/Co<br>$m_T = 1.87(4)$ $\mu_B$/Co<br>T=15 K **FM1** | $m_x = 1.49(3)$ $\mu_B$/Co<br>$m_z = 0$<br>T=140 K **FM2** |
| Magnetic structure | 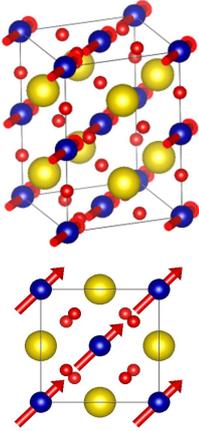 | 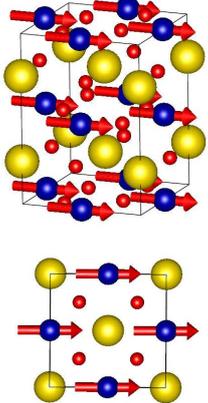 |